\documentclass{aastex631}

\def\gcc{\hbox{\rm\hskip.35em  g cm}$^{-3}$}
\def\cms{\hbox{\rm\hskip.35em  cm s}$^{-1}$}

\def\lap{\hbox{${_{\displaystyle<}\atop^{\displaystyle\sim}}$}} 
\def\gap{\hbox{${_{\displaystyle>}\atop^{\displaystyle\sim}}$}}

\DeclareMathAlphabet\mathbfcal{OMS}{cmsy}{b}{n}

\newcommand{\half}{\frac{1}{2}}

\newcommand{\ubf}         {\mathbf u}

\newcommand{\nbf}         {\mathbf n}

\newcommand{\dlbf}         {\mathbf dl}

\newcommand{\kappabf}         {\mbox{\boldmath$\kappa$}}
\newcommand{\Fbf}         {\mathbf F}

\newcommand{\fbf}         {\mathbf f}

\newcommand{\vbf}         {\mathbf v}

\newcommand{\rbf}         {\mathbf r}

\newcommand{\Omegabf}         {\mbox{\boldmath$\Omega$}}
\newcommand{\omegabf}         {\mbox{\boldmath$\omega$}}

\def\half{\frac{1}{2}}

\newcommand{\be}{\begin{equation}}
\newcommand{\ee}{\end{equation}}

\usepackage{movie15}
\usepackage{graphicx}
\usepackage{float}
\usepackage{hyperref}
\usepackage{xcolor}
\usepackage{amsmath}

\begin{document}

\title{Superfluid Rivers in Spinning-down Neutron Stars}

\author{Yuri Levin}
\affiliation{Physics Department and Columbia Astrophysics Laboratory,
Columbia University, 538  West 120th Street New York, NY 10027, USA}
\affiliation{Center for Computational Astrophysics, Flatiron
  Institute, 162 Fifth Avenue, New York, NY 10010, USA}
\affiliation{Department of Physics and Astronomy, Monash University,
  Clayton
  VIC 3800, AUSTRALIA}
\author{Bennett Link}
\affiliation{Department of Physics, Montana State University, Bozeman,
MT 59717, USA}

\begin{abstract}
  We study the motion of neutron superfluid vortices in a
  spinning-down neutron star, assuming axisymmetry of the flow and
  ignoring motion of vortices about the rotation axis.  We find that
  the vortex array, if initially rectilinear, is soon substantially
  deformed as the star spins down; vortices are swept outward by the
  Magnus force, accumulating in regions of the inner crust where they
  pin, accompanied by significant bending of the vortex array.  As the
  star spins down to below a spin rate of $\sim 20$ Hz (twice the spin
  rate of the Vela pulsar), the Magnus and pinning forces gradually
  compress the vortex array into dense sheets that follow spherical
  shells. In some cases, the vortex array bends on itself and
  reconnects, forming one or more tori of vortex rings that contain superfluid
  ``rivers" with significant angular momentum. Vortex sheets
  are likely to form near the base of the inner crust, in the regime
  of nuclear pasta.  
\end{abstract}


\section{Introduction}

The manner in which the neutron superfluid (nSF) spins down in a
neutron star (NS) is likely related to various observed spin
irregularities.  Spin glitches in NSs are generally attributed to
sudden transitions, driven by stellar spin-down, between different
metastable states of pinned vorticity in the inner crust \citep{ai75}
or the outer core (\citealt{rzc98}; see \citealt{haskell2015models} for a
review). Pulsars also exhibit long-timescale stochastic spin
irregularities, known as {\sl timing noise} (see
\citealt{hobbs_etal10} and references therein). Timing noise could
arise from a number of superfluid-related processes such as turbulence
\citep{greenstein1970superfluid,ml14,lasky2015pulsar}, stochastic
transport of vorticity \citep{alpar_etal85,jones90b}, vortex
avalanches \citep{cheng_etal88,wm08}, and magnetospheric processes
unrelated to superfluid dynamics
\citep{cheng87a,cheng87b,ulw06,lyne_etal2010}.

Knowledge of the structure of the superfluid flow - in particular,
columnar and laminar versus turbulent - is essential for the
development of realistic models of NS spin-down and timing
irregularities.  \cite{rs74} argued long ago that the array of
superfluid vortices in a spinning-down NS is effectively very rigid
and so  remains rectilinear despite the presence of forces on vortices
that vary with position in the star; they concluded that the vortex
array will evolve without significant bending or tangling - a
superfluid version of the Taylor-Proudman theorem of classical
hydrodynamics - precluding the development of superfluid turbulence.
As a result of the work of \cite{rs74}, and also perhaps as an
understandable simplification of a complex problem, much work on the
hydrodynamics of the NS interior has assumed that the vortex array
remains rectilinear as the NS evolves (e.g.,
\citealt{alpar_etal84a,alpar_etal85,le96,ll02,antonelli2017axially}).
Other work has assumed that the nSF finds itself in a turbulent state,
and has explored the consequences of this assumption (e.g.,
\citealt{greenstein1970superfluid,andersson_etal07,lasky2013stochastic,ml14,haskell2020turbulent}).
A number of interesting simulations of superfluid turbulence in NSs
introduce laminar-turbulent transitions by hand (e.g.,
\citealt{peralta_etal06}).

In this paper we address the evolution of the superfluid vortex array
in axisymmetry, accounting for vortex pinning in the inner crust.  We find
that for realistic vortex pinning, the Taylor-Proudman theorem is
strongly violated. Instead, in regions where the pinning strength has
strong gradients, vortices accumulate into dense spherical sheets. For
some distributions of the pinning force, the spherical sheets bend and
reconnect to make tori with with  superfluid rivers flowing inside them.
Generally, we suspect that the vortex array evolves into complex
configurations that could become tangled, with an accompanying
superfluid flow that could be turbulent.

The plan of this paper is as follows. In \S \ref{sd} we derive the
equations of motion for superfluid vorticity.  In \S \ref{tp}, we
derive a superfluid version of  Taylor-Proudman Theorem, and show that the theorem is
typically violated when pinning forces are present. In \S \ref{sv} we
describe the motion of a single vortex.  In \S \ref{gd} we describe
the global dynamics in axisymmetry.  In \S \ref{gm} we present a
simple global model for vortex mobility that is easily solved.  In \S
\ref{sims} we present our simulations. In \S \ref{discussion} we
discuss our results.

\section{Superfluid Dynamics}
\label{sd}

A superfluid rotates by establishing an array of quantized  vortices of
microscopic cross section.  The circulation about a vortex is fixed
according to
\begin{equation}
  \oint\vbf_s(\rbf,t)\cdot\dlbf=\kappa,
\label{vort_cons}
\end{equation}
where $\vbf_s(\rbf,t)$ is the superfluid velocity at location $\rbf$
at time $t$; $\kappa=2\pi\hbar/m$ is the circulation where $m$ is the
mass of the fundamental boson in the system, twice the neutron mass in
the case of a neutron superfluid.  An important distinction between a
superfluid and a classical fluid is that vortices in a superfluid are
physical entities that can move at their own velocity distinct from
that of the superfluid, interact with their environment, and pin or
experience physical drag from other parts of the system. Regardless of
 how superfluid vortices move,  each vortex always carries a
quantum of circulation according to eq. \ref{vort_cons}. 

Consider an incompressible superfluid. The momentum equation in the
inertial frame is 
\begin{equation}
  \frac{\partial\vbf_s}{\partial
    t}+\vbf_s\cdot\nabla\vbf_s=-\nabla\mu-\nabla\phi+\rho_s^{-1}\fbf,
\end{equation}
where $\mu$ is the chemical potential, $\phi$ is the gravitational
potential, $\fbf$ is the force per unit volume acting on the
superfluid from vortex pinning, mutual friction, and vortex tension, and $\rho_s$ is the mass density of the
superfluid. It is convenient to work in the frame corotating with the
crust at angular velocity $\Omegabf(t)$ and we denote quantities in
this frame with the superscript *.  The velocity is
  \begin{equation}
    \vbf^*_s(\rbf,t)=\vbf_s(\rbf,t)-\Omegabf\times\rbf.
  \end{equation}
In the rotating frame, the momentum equation is
\begin{equation}
  \frac{\partial\vbf^*_s}{\partial
    t}+\vbf^*_s\cdot\nabla\vbf^*_s
  =-\nabla\left(\mu+\phi-\half(\Omegabf\times\rbf)^2\right)
-2\Omegabf\times\vbf^*_s
  -\frac{d\Omegabf}{dt}\times\rbf
  +\rho_s^{-1}\fbf.
\label{momentum_star}
\end{equation}
Taking the curl, and using a vector
identity\footnote{$\half\nabla
  v^2=\vbf\times\omegabf+\vbf\cdot\nabla\vbf$.}, gives
\begin{equation}
  \frac{\partial\omegabf}{\partial
    t}=\nabla\times\left(\vbf^*_s\times\omegabf+\rho_s^{-1}\fbf\right).
\label{domdt0}
\end{equation}
Here $\omegabf$ is the total vorticity measured in the inertial frame
\begin{equation}
  \omegabf=\nabla\times\vbf_s=2\Omegabf+\nabla\times\vbf^*_s=2\Omegabf+\omegabf^*,
\end{equation}
and $\omegabf^*\equiv\nabla\times\vbf^*_s$ is the vorticity in the
rotating frame.  In a quantum fluid, vortices are persistent
structures to which we can assign a local velocity $\vbf^*_v$ in the
rotating frame.  Vortex motion obeys the conservation law,
\begin{equation}
  \frac{\partial\omegabf}{\partial
    t}=\nabla\times\left(\vbf^*_v\times\omegabf\right). 
\label{domdt1}
\end{equation}
The force $\fbf$ contains contributions from mutual friction  and bending forces
of the vortex array.  Comparing eqs.  \ref{domdt0} and \ref{domdt1}
gives the vortex equation of motion
\begin{equation}
\rho_s^{-1}\fbf=  (\vbf^*_v-\vbf^*_s)\times\omegabf,
\label{vortex_eom1}
 \end{equation}
up to the gradient of an arbitrary scalar potential, which we absorb into the
definition of $\mu$.
The right-hand side is the Magnus force (per unit mass). 
For $\fbf=0$, the vortex lines move with the superfluid, while
 for $\fbf\ne 0$, vortices are forced to move locally at a different
 velocity $\vbf^*_v\ne\vbf^*_s$.

 We can write eq. \ref{vortex_eom1} as an
 equation of motion for a single vortex by defining the areal density
 of vortices $n_v$ related to the vorticity through
\begin{equation}
  n_v\kappabf=\nabla\times {\bf v}_s\equiv\omegabf,
\label{kappa_def}
\end{equation}
where $\kappabf$ has magnitude $\kappa$ in the direction of the local
vorticity $\omegabf$. Noting that
the force per unit length on a single vortex is $\Fbf_v=n_v^{-1}\fbf$,
gives
\begin{equation}
  \rho_s\kappabf\times(\vbf^*_v-\vbf^*_s)+\Fbf_v=0.
\label{vortex_eom2}
\end{equation}

\section{Superfluid Taylor-Proudman Theorem}
\label{tp}

For steady rotation of a classical fluid in the limit that the
Coriolis force is much larger than the inertial force, the fluid flow
does not depend on position along the axis of rotation. If a solid
body is slowly pulled through the fluid, the fluid must flow around
the body, and  a vertical Taylor column is established along the
axis of rotation.  In a rotating superfluid the analogous effect is
seen experimentally as vortices tend to become straight and align with
the rotation axis in a rapidly rotating vessel.  We review the
derivation of the Taylor-Proudman for a rotating fluid.  We then show that the theorem
is violated in the presence of pinning forces in a quantum fluid. 

In a NS, the spin-down torque acts on the crust and is communicated
to the nSF  over a long time scale $t_{\rm sd}$, so that
$\Omega t_{\rm sd} \ll 1$.  We therefore consider very slow variations
in $\omegabf$ when a spatially-dependent force $\fbf$ is
present. Eq. \ref{domdt1} gives
\begin{equation}
  (\omegabf\cdot\nabla)\vbf^*_s-(\vbf^*_s\cdot\nabla)\omegabf=-\nabla\times(\rho_s^{-1}\fbf).
\label{vort2}
\end{equation}
The Taylor-Proudman theorem applies when the Coriolis force dominates
both the inertial force in eq. \ref{momentum_star} and
$\rho_s^{-1}\fbf$. In terms of vorticity, the Coriolis force dominates
the inertial force when 
$2\Omega\gg |\omegabf^*|$.  In this limit, with
$\Omegabf=\hat{z}\Omega$ where $\hat{z}$ is a unit vector along
the rotation axis, eq. \ref{vort2} becomes 
\begin{equation}
  \frac{\partial\vbf^*_s}{\partial z}=-\frac{1}{2\Omega}\,\nabla\times(\rho_s^{-1}\fbf).
\label{ss}
\end{equation}
Typically in a classical fluid $\rho_s^{-1}\fbf$ is determined by the
properties of, for example, an obstacle affecting the flow. In the
limit of fast rotation the right-hand side goes to zero and we arrive
at the Taylor-Proudman theorem; all three components of $\vbf^*_s$ become
independent of $z$. For a quantum fluid $\rho_s^{-1}\fbf$ might be
due to mutual friction, which scales as the vortex density and hence
as $2\Omega$ in the limit of fast rotation; then
$\partial\vbf^*_s/\partial z$ does not tend to zero as in a
classical fluid.  In cylindrical coordinates $(\varrho,\phi^*,z)$, and
assuming axisymmetry, eq. \ref{domdt0} becomes in the limit of fast
rotation
\begin{align}
  & \frac{\partial\omega_{\varrho}}{\partial t}=
    \frac{\partial}{\partial t}\left(\frac{\partial v^*_{\phi^*}}{\partial z}\right)=
    -2\Omega\frac{\partial v^*_{\varrho}}{\partial z}
    +\frac{\partial \tilde{f}_{\phi^*}}{\partial z}\\
  & \frac{\partial\omega_{\phi^*}}{\partial
    t}=\frac{\partial}{\partial t}\left(\frac{\partial{v^*_{\varrho}}}{\partial z}
    -\frac{\partial v^*_z}{\partial\varrho}\right)=
    2\Omega\frac{\partial v^*_{\phi^*}}{\partial z}
    +\frac{\partial\tilde{f}_{\varrho}}{\partial z}
    -\frac{\partial\tilde{f}_{z}}{\partial\varrho}
    \label{middle} \\
& \frac{\partial\omega_{z}}{\partial t}= \frac{\partial}{\partial
  t}\left(\frac{1}{\varrho}\frac{\partial}{\partial\varrho}(\varrho v^*_{\phi^*})\right)
  =
  2\Omega\frac{\partial v_{z}}{\partial z}
  +\frac{1}{\varrho}\frac{\partial}{\partial\varrho}(\varrho\tilde{f}_{\phi^*}), 
  \end{align}
  where $\tilde{\fbf}\equiv\rho_s^{-1}\fbf$ and all quantities are
  functions of the coordinates $(\varrho,z)$.  If $f_{\varrho}=f_z=0$,
  $v^*_{\phi^*}$ is independent of $z$ implying that the vortex
  array remains rectilinear as it moves \citep{rs74}; the array does
  not bend.  Generally, though, $f_{\varrho}\ne 0$, and $v^*_{\phi^*}$
  becomes dependent upon $z$; the array bends as it moves.  The
  final term in eq. \ref{middle} becomes non-zero  as the array bends. From
  eq. \ref{middle}, we can estimate the angle by which the vortex
  lattice deviates locally from straight:
  \begin{equation}
    \delta\theta\sim\frac{v^*_{\phi^*}}{\varrho\Omega}\sim\frac{\rho_s^{-1}f_{\varrho}}{2\varrho\Omega^2}.
    \end{equation}
The deformation of the vortex lattice scales as $\Omega^{-1}$ for a
force that is proportional to the vortex density.  Conversely, lattice
deformations can be large in the limit of slow rotation, as the
simulations we present later in this paper show. 

It is instructive to consider the Taylor-Proudman limit for a quantum
fluid in terms of vortex motion.  The vortex motion in the steady
state satisfies, from eq. \ref{domdt1}
\begin{equation}
  (\omegabf\cdot\nabla)\vbf^*_v-\omegabf(\nabla\cdot\vbf^*_v)
  -(\vbf^*_v\cdot\nabla)\omegabf=0.
\label{vortex_ss}
\end{equation}
In the limit of fast rotation with $2\Omega\gg|\omegabf^*|$,
\begin{equation}
  2\Omega\left(\frac{\partial\vbf^*_v}{\partial
      z}-\hat{z}\,\nabla\cdot\vbf^*_v\right)\rightarrow 0.
  \end{equation}
In steady state, the vortex motion becomes independent of the axial
coordinate $z$, the lateral vortex velocity is nondivergent and
independent of the axial coordinate, and the axial velocity decouples
from the lateral velocity (see also
\citealt{holm2001renormalized}).  Hence, in this steady-state limit of
fast rotation, the vortices straighten and become parallel to the
rotation axis. 

The Taylor-Proudman theorem generally does not
hold if vortices are immobilized in parts of the fluid volume (as
opposed to only at boundaries) by pinning forces.  In this case the
pinning force follows from eq. \ref{vortex_eom1} with $\vbf^*_v=0$:
\begin{equation}
  \rho_s^{-1}\fbf=\omegabf\times\vbf^*_s.
\end{equation}
Now $\partial\omegabf/\partial t=0$ because the vortices are
immobilized and the vortices cannot adjust their configuration to
make $\vbf^*_s$ independent of $z$ for an arbitrary pinning force.  
If $\rho_s^{-1}\fbf_{\varrho}$ varies along $z$, for
example, so will $\vbf^*_{\phi^*}$, implying that the vortex array is
not rectilinear.  This is the situation we explore further in this
paper. 

Finally, we note that the force $\fbf$ contains a contribution from rigidity of the vortex
lattice, which is typically tiny compared to pinning forces; see next
section. 

\section{Motion of a Single Vortex}
\label{sv}

Let the vortex shape, measured with respect to the $z$ axis, be given by
the two-dimensional displacement vector
$\ubf(z,t)=\hat{x}u_x+\hat{y}u_y$, so the vortex velocity is
$\vbf_v^*=\partial\ubf/\partial t$. For small
bending angles, so that $|\partial\ubf/\partial z| \ll 1$, eq. \ref{vortex_eom2}
can be written (see, also, \citealt{schwarz1977theory,schwarz1978turbulence})
\begin{equation}
\rho_s\kappabf\times\left(\frac{\partial\ubf}{\partial
    t}-\vbf^*_s\right)+ T_v\frac{\partial^2\ubf}{\partial  z^2}+\Fbf_L=0. 
\label{vortex_eom0}
\end{equation}
The first term is
the Magnus force. The second term is the force (per unit length) from
bending the vortex; $T_v$ is the vortex tension. The last term $\Fbf_L$ is the force due
to interaction of the vortex with the nuclear lattice. The second and
third terms are $\Fbf_v$ appearing in eq. \ref{vortex_eom2}. 
The tension is typically
\begin{equation}
  T_v\simeq \frac{\rho_s\kappa^2}{4\pi}\Lambda, 
\end{equation}
where $\Lambda$ is a factor of 3-10; $T_v \sim 1$ MeV fm.  As
explained below, vortex tension is important against pinning forces
only over length scales below $\sim 10^3$ fm. 

The force of the lattice on the vortex has two contributions: (i) a
static, conservative contribution and (ii) a velocity-dependent,
damping force due to the excitation of lattice phonons by the moving
vortex that carry energy away from the vortex \citep{eb92,jones92}. To
treat these two contributions, we take the simple form
  \begin{equation}
    \Fbf_L=-\nabla_\perp V -\eta\vbf_v^*,
\label{fL}
  \end{equation}
where $\vbf^*_c$ is the velocity of the crust. 
  The first term gives the static force; $V$ is the potential, 
  and the gradient is  perpendicular to the vortex. The second term is
  the drag force, and $\eta$ is the drag coefficient.
  A non-dissipative, velocity-dependent contribution to $\fbf_L$ is
  possible, but shall not be considered here; see \citet{swc99}.

Numerical solutions of eqs. \ref{vortex_eom0} and \ref{fL} for various
approximations to the nuclear lattice show that the vortex often responds
with a slip-stick character to an applied force
$\rho_s\kappabf\times\vbf^*_s$ \citep{link_levin22}. For zero applied
force, the vortex settles into a pinned configuration whose solution is
given by
\begin{equation}
T_v\frac{\partial^2\ubf}{\partial  z^2}-\nabla_\perp V=0.
\end{equation}
The response of a pinned vortex to a slowly increasing applied force
depends on the properties of the lattice, its orientation, whether the
vortex-lattice interaction is attractive or repulsive, and the drag
coefficient $\eta$.  We now summarize the characteristics of vortex
motion in the case of low drag, in the sense
$\eta/\rho_s\kappa\ll 1$. If the lattice is regular and repulsive,
for most lattice orientations the vortex is able to move through the
propagation of kinks.  As the force becomes large, the vortex enters a
ballistic regime in which the conservative contribution to the lattice
force and the tension contribution average nearly to zero.  In this
limit, the vortex motion is given by
\begin{equation}
  \rho_s\kappabf\times\vbf^*_v-\eta\vbf^*_v=0.
\label{ballistic}
\end{equation}
For some lattice orientations, the kinks hang up and the vortex
remains pinned.  As the force is increased further  for this case, the vortex unpins
suddenly at a critical value of the force, with Kelvin waves
propagating along its length. If the force is then reduced, the vortex
eventually repins quite suddenly, but typically at a value of the
force that is smaller than the value of the force at which the vortex
unpinned.  This hysteretic attribute of vortex motion is due to the
fact that the vortex is an extended object that supports internal
vibrational degrees of freedom.  Impurities in the lattice tend to
prevent the propagation of kinks, giving a well-defined transition
from pinned to ballistic at a critical value of the force.  If the
lattice is regular and attractive, there is a well-defined transition
from pinned to ballistic for most lattice orientations.  If the
lattice is strongly disordered, or if the pinning sites are randomly positioned as in a glass, then a well-defined transition from pinned to ballistic
occurs whether the lattice is attractive or repulsive.  Typical values
for the critical force are $10^{16}$ dyn cm$^{-1}$ to
$10^{17}$ dyn cm$^{-1}$, corresponding to critical superfluid
velocities of $v_{\rm crit}=\vert \vbf^*_s\vert$ of $10^{-5}c$ to
$10^{-4}c$.

The motion in the ballistic regime is given by solution to eq. \ref{ballistic}:
\begin{equation}
  \vbf^*_v=\cos\theta\,(\vbf^*_s\cos\theta
    +[\vbf^*_s\times\hat{\kappa}]\,\sin\theta)
  \qquad \tan\theta\equiv\frac{\eta}{\rho_s\kappa}.
\label{solution_ballistic}
\end{equation}
Locally, the vortices move at an angle $\theta$ with respect to the vector
$\vbf^*_s$; $\theta$ is sometimes referred to as the ``dissipation
angle''.  In the limit of low drag, $\eta/\rho_s\kappa <<1$ and
$\theta\simeq\eta/\rho_s\kappa$, giving
\begin{equation}
  \vbf^*_v\simeq \vbf^*_s+  (\vbf^*_s\times\hat{\kappa})\,\theta.
\label{lowdrag}
\end{equation}
The vortex moves nearly with the superfluid, with slow drift along
$\vbf^*_s\times\hat{\kappa}$.  In the limit of high drag, $\eta/\rho_s\kappa >>1$ and
$\theta\rightarrow \pi/2$, giving
\begin{equation}
  \vbf^*_v\simeq \left(\frac{\rho_s\kappa}{\eta}\right)\vbf^*_s\times\hat{\kappa}.
\label{highdrag}
\end{equation}

The solutions described above apply for low drag. 
In the limit of high drag, an unpinned vortex moves nearly with the
crust, with slow drift along $\vbf^*_s\times\hat{\kappa}$.  We refer
to this state of unpinned vortex motion as ``vortex drift''  to
distinguish the motion from that of the low-drag, ballistic regime. 

For vortex motion in the inner crust, the drag coefficient is typically
$\gamma\equiv\eta/\rho_s\kappa\sim 10^{-3}$ \citep{eb92,jones92}, and depends
on velocity.\footnote{For
low-velocity motion ($v^*_s<10^3$ \cms), \citet{jones90a} finds
$\gamma\sim 10^{-5}-10^{-4}$. That calculation assumes that
vortices cannot pin, while \citet{link_levin22} find that vortices pin strongly in this
regime.}
In the outer core, SFn vortices are predicted to acquire magnetization due to
entrainment of protons by the neutrons that circulate around the
vortices \citep{als84}.  Electrons scatter with the magnetic moments
of the SFn vortices. The relaxation time for relative motion
between the neutrons and electron-proton plasma is \citep{als84}
\begin{equation}
  \tau_v \sim P(\mbox{s})\mbox{ s},
\end{equation}
where $P(\mbox{s})$ is the spin period of the system in seconds.  The
dissipation angle is related to the relaxation time at spin rate
frequency $\Omega$ through \citep{link14_slippage}
  \begin{equation}
    \sin 2\theta =(\Omega\tau_v)^{-1}\longrightarrow \sin\theta\simeq (4\pi)^{-1}.
  \end{equation}
 According to this estimate,  vortex motion in the core is in
the low-drag regime. 
However, neutron vortices are also expected to pin to the far more numerous flux
tubes  of the outer core (see, e.g., \citealt{srinivasan_etal90,rzc98}; \citealt{jones91}),
 restricting the mobility of the neutron vortices;  in order to move, the
vortices must cut through flux tubes; the dissipation associated 
with this process is poorly understood.

The bending force per unit length - the first term in
eq. \ref{vortex_eom0} - is $\sim T_v/R_v$, where $R_v$ is the radius
of curvature of a bent vortex.  For a pinning force of $10^{17}$ dyn
cm$^{-1}$ $\simeq 10^{-3}$ MeV fm$^{-2}$, the pinning force will
dominate the bending force for $R_v\gap 10^3$ fm = $10^{-10}$
cm. Vortex tension is therefore completely negligible for the problem
of macroscopic fluid flow in a NS, though it plays an essential
role in pinning and slip-stick dynamics, as described
in more detail in \citet{link_levin22}. 

\section{Global Dynamics}
\label{gd}

Consider a vortex that is initially pinned in a spinning-down NS with
a nSF flow velocity $\vbf_s^*$ in the frame of the crust.  As the
stellar crust spins down, the superfluid flow speed past the pinned
vortex will approach the  local critical value for unpinning. At
zero temperature, the vortex unpins only when the flow speed becomes
critical.  For low-drag vortex motion, the vortex moves nearly with
$\vbf_s^*$ when the vortex is unpinned, while for high-drag vortex motion,
the vortex drifts slowly transverse to the
$\vbf_s^*$ (along $\vbf_s^*\times\hat{\kappa}$; see
eq. \ref{highdrag}). 

At finite temperature, thermal excitations cause a pinned vortex to
slowly ``creep'' when the flow velocity is slightly subcritical
\citep{alpar_etal84a, leb93, sa09, link14_slippage}.  The dissipation
angle for motion through creep can differ from the value given by
$\tan\theta=\eta/{\rho_s\kappa}$ in the ballistic regime.  As we
show in a forthcoming publication \citep{link_levin2023}, for low-drag thermal creep the
vortex has comparable velocity components along and transverse to
$\vbf_s$, while for high-drag thermal creep the vortex motion is nearly
transverse to $\vbf_s^*$, as for zero temperature.

If vortex motion has a significant component along $\vbf_s^*$, the
vortex array will become substantially twisted as the NS spins down.
For both illustration and simplicity, we henceforth neglect vortex
motion along $\vbf_s^*$, and consider axisymmetric superfluid flow.
We assume that, in the rotating frame, the vortices are always
poloidally-directed.  This condition is rigorously satisfied in the
limit of high drag, but is not satisfied in the limit of low drag if
stellar spin down causes the vortex array to make transitions between
a pinned state and a ballistic state.  The limitations of our
treatment are discussed \S \ref{discussion}.

We use cylindrical coordinates in the inertial and rotating frames
with common origin and $z$ axes that coincide:
\begin{align}
  \rbf &=\varrho\hat{\varrho}+z\hat{z} \\
  \rbf &=\varrho\hat{\varrho}^*+z\hat{z}.
  \end{align}
The rotation vector
of the crust $\Omegabf$ is along $\hat{z}$. It is convenient to work with the total
circulation, defined as
\begin{equation}
    \Phi(\varrho,z)\equiv 2\pi \varrho v_s(\varrho,z).
\label{Phi-def}
\end{equation}
The quantity $\Phi/\kappa$ is equal to the total number of vortices
enclosed by a horizontal circle of radius $\varrho$ at height $z$.

We now establish a theorem for the time evolution of the circulation:
{\em so long as $v^*_s$ exceeds the local threshold for unpinning, the
  Magnus force will drive a local reduction in the circulation.} 
Eqs. \ref{domdt1} and \ref{vortex_eom1}  give, by Stokes's Theorem,
\begin{equation}
  \frac{\partial}{\partial t}\Phi(\varrho,z,t)=-2\pi
  \varrho\,\vbf^*_v\cdot(\hat{\phi}^*\times\omegabf)=2\pi\varrho\,\rho_s^{-1}f_{\phi^*}.
\label{dcircdt}
\end{equation}
The rate of change of the circulation is equal to the flux of the
total vorticity vector $\omegabf$ through the horizontal circle
centered on the axis of symmetry, whose points have fixed poloidal
coordinates $(\varrho,z)$. If the superfluid rotates faster than the
crust ({\sl viz}., $v_s^*>0$), the azimuthal component of drag force,
which opposes the azimuthal motion of vortices, is along
$-\hat{\phi^*}$, so $f_{\phi^*}$ is negative.  From eq. \ref{dcircdt},
the Magnus force then drives a local reduction of the circulation;
this remains valid even if the vortices are directed opposite to the
 direction of stellar rotation. 

\section{Global Model for Vortex Mobility}
\label{gm}

Based on eq. \ref{dcircdt}, we consider a simple model for the
mobility of vortices: the vortex is mobile only when the Magnus force
exceeds a certain position-dependent critical value, that is, when
$v^*_s$ exceeds $v_{\rm crit}(\rbf)$. The critical velocity
$v_{\rm crit}$ is a function of the mass density, which has a
spherically-symmetric distribution, so $v_{\rm crit}$ is a function of
the spherical radius $r$.   When this unpinning threshold is exceeded,
vortex motion reduces the circulation locally.   Returning to
the inertial frame, where the crust velocity is
$\varrho\Omega(t)\hat{\phi}$ and the superfluid velcocity is 
$\vbf_s=(\varrho\Omega(t)+v_s^*)\hat{\phi}$, we have the following
prescription for how the  speed  $v_s$ evolves with time:
\begin{itemize}
    \item 
So long $v_s(\varrho,z)<\varrho\Omega(t)+v_{\rm crit}(r)$, the
constant-circulation surface is pinned ; $\Phi(\varrho,z)$  and 
$v_s(\varrho,z)$ remain fixed. 

\item Once $v_s(\varrho,z)\ge \varrho\Omega(t)+v_{\rm crit}(r)$, the
  constant-circulation surface (representing vortices) becomes
  unpinned and moves under the action of the Magnus force. This motion
  reduces $v_s(\varrho,z)$. We assume that the timescale for the
  unbalanced vortex motion is much shorter than that of the pulsar
  spindown; in that case $v_s$ reduces  to the value very close to
  $\varrho\Omega(t) +v_{\rm crit}(r)$ and the vortices/constant circulation
  surfaces become pinned again.

  \medskip

\end{itemize}

We consider steady spin-down of the crust, and therefore once
$v_s(\varrho, z)$ reaches $\varrho\Omega+v_{\rm crit}(r)$, it will
track this quantity as $\Omega$ further decreases. Hence, 
\begin{equation}
    v_s(\varrho,z,t)=\min\left[\varrho\Omega_0, \varrho\Omega(t)+v_{\rm crit}(r)\right].
\label{solution}
\end{equation}
Here $\Omega_0$ is the initial angular velocity of the star, and it is
assumed that the superfluid was initially co-rotating with the star.
The circulation is
\begin{equation}
  \Phi(\varrho,z,t)=2\pi\varrho\,\min\left[\varrho\Omega_0,
    \varrho\Omega(t)+v_{\rm crit}(r)\right].
\end{equation}
The circulation is straightforward to calculate.  The contours of
constant circulation trace out vortices. 

The remarkable feature of eq. \ref{solution} is that the superfluid
velocity is determined locally, irrespective of the history of the
pulsar spindown and of the pinning strengths at other location. This
feature is a consequence of the assumed  axisymmetry with negligible
vortex motion around the rotation axis.  The neighbouring
constant circulation surfaces do not feel attraction or repulsion to
each other; the motion of one does not change the superfluid velocity
at the other. One immediate corollary is that the strength of pinning
in the core does not affect the motion of nSF vortices in the
crust.

In the initial stages of NS spin down, the star is spinning down
quickly and $\varrho\Omega_0>\varrho\Omega(t)+v_{\rm crit}(r)$ everywhere in
star except very close to the rotation axis. Then
\begin{equation}
  v_s(\varrho,z,t)\rightarrow\varrho\Omega(t)+v_{\rm crit}(r) .
  \end{equation}
The vorticity vector can be expressed as
$\omegabf=\hat{\kappa}n_v\equiv\kappa\nbf_v$, where $\nbf_v$ is the
vector areal density of the vortex array. 
The vortex density  $\nbf_v$ at location ($r$, $\theta$) in
spherical coordinates is given by (except close to the rotation axis)
\begin{equation}
  \kappa\nbf_v\equiv\hat{\kappa}n_v=\nabla\times\vbf_s=
  \hat{r}\left[2\Omega(t)\cos\theta+\frac{\cot\theta}{r}v_{\rm
      crit}(r)\right]
  -\hat{\theta}\left[2\Omega(t)\sin\theta+\frac{1}{r}\frac{d}{dr}\left\{rv_{\rm crit}(r)\right\}\right].
\label{vector_density}
  \end{equation}
As described in the next section, radial gradients in the pinning
force can be large in the inner crust.  In such regions (except close
to the rotation axis) 
\begin{equation}
  \kappa\nbf_v\simeq
  \hat{z} 2\Omega(t) - \hat{\theta} \frac{1}{r}\frac{d}{dr}\left\{rv_{\rm crit}(r)\right\}.
  \end{equation}
At high spin rate,  $\nbf_v$ is uniform and points in the $z$
direction.  As $\Omega(t)$ becomes sufficiently small,  the vorticity
vector begins to point along $\pm\hat{\theta}$ in regions where the
radial gradient in the pinning force is large.  This transition begins
when $2\Omega(t)\lap r^{-1}d\{rv_{\rm crit}(r)\}/dr$,
corresponding to a spin period of 
\begin{equation}
  P \gap \frac{4\pi\sigma_p}{v_{\rm crit}^{\rm max}}\simeq 0.2\,
  \left(\frac{\sigma_p}{0.05R}\right)
  \left(\frac{v_{\rm crit}^{\rm max}}{3\times 10^6\,
      \mbox{\cms}}\right)^{-1} \mbox{ s}.
\end{equation}
where $\sigma_p$ is the characteristic dimension over which the
pinning forces varies, $0.05R$ is a fiducial value for the thickness
of the crust, and $v_{\rm crit}^{\rm max}$ is the maximum value of the
critical velocity for pinning.  Hence, as the star spins down the
vorticity becomes compressed into dense sheets along
$\pm\hat{\theta}$, even if the vortex array is uniform initially; a
sheet will be along $-\hat{\theta}$ in a region with
$d(rv_{\rm crit})/dr>0$ and along $\hat{\theta}$ in a region
with $d(rv_{\rm crit})/dr<0$.

\section{Pinning Models and Simulations}
\label{sims}

There has been substantial disagreement over the magnitude and sign of
the vortex--nucleus interaction in the inner crust.  Quantum
calculations (using a mean-field Hartree--Fock--Bogoliubov
formulation) show a repulsive interaction with an energy of up to
$\sim 3$ MeV \citep{abbv07,avogadro2008vortex} throughout the inner
crust, while semiclassical calculations (using a local density
approximation) show a repulsive interaction of $\sim 1-2$ MeV below an
average baryon density of $\sim 10^{-2}$ fm$^{-3}$
($\sim 2\times 10^{13}$ \gcc) which turns attractive with a strength
of $\sim 5 $ MeV at higher densities
\citep{pvb97,donati_pizzochero2003,dp2004,dp06}. A step toward
resolving the controversy was made by \citet{wlazlowski_etal16}, using
density-functional theory (in principle, an exact approach); they find
that the vortex--nucleus interaction is always repulsive in the
average baryon density range 0.02 fm$^{-3}$ ($3\times 10^{13}$ \gcc)
to 0.04 fm$^{-3}$ ($7\times 10^{13}$ \gcc), with a force of $\sim 1$
MeV fm$^{-1}$ over a range of $\sim 4$ fm, corresponding to an
interaction energy per nucleus of $\sim 4 $ MeV.  The strength of
pinning also depends on the symmetry of the nuclear lattice
(\cite{link_levin22}; recall discussion of \S \ref{sv}).  Pinning is
relatively strong in a regular, attractive lattice.  Pinning occurs
less readily in a regular, repulsive lattice, with no sharp transition
to pinning for many lattice orientations.  The presence of impurities
in an otherwise regular lattice, such as dislocations, nuclei with
different charge than their neighbors, or mono-vacancies (missing
nuclei), generally enhance pinning.  Pinning in nuclear glass is
generally strong.  Despite these uncertainties, it is clear that the
basic energy scale in the pinning interaction is $\sim 1$ MeV, with
spatial variations over length scales much shorter than the crust
thickness of $\sim 0.05$ the stellar radius.  Vortices are also
predicted to pin to the far more numerous flux tubes of the outer core
(see, e.g., \citealt{srinivasan_etal90,rzc98}; \citealt{jones91}).

Though the strength and distribution of the pinning force is
uncertain, we can proceed with two illustrative
examples for the distribution of the pinning force from which we can
draw some general conclusions.  In both cases we
make the pinning region deeper in the star and somewhat thicker than
in a real NS crust for the sake of visibility. This choice has little
effect on our chief results.

{\sl Example Model 1: spherically-symmetric Gaussian. }
\begin{equation}
  v_{\rm crit}(r)=v_{\rm crit}^{\rm max}\,{\rm e}^{-(r-r_0)^2/2\sigma^2},
\end{equation}
where $r=\sqrt{\varrho^2+z^2}$ is the spherical radius coordinate, $r_0$ is the coordinate
at which $v_{\rm crit}$ takes its largest value, and $\sigma$ is the
width of the distribution.

{\sl Example Model 2: spherically-symmetric Gaussian with sinusoidal
oscillations.}  The purpose of this model is to show the effects of  large spatial
gradients in pinning strength as indicated by  the calculations of
\cite{pvb97} (see, also, \citealt{donati_pizzochero2003,dp2004,dp06}). 
\begin{equation}
  v_{\rm crit}(r)=v_{\rm crit}^{\rm max}\,{\rm e}^{-(r-r_0)^2/2\sigma^2}(1-\sin kr),
\end{equation}
where $k$ is the wave number of spatial oscillations. 

We fix $v_{\rm crit}^{\rm max}=10^{-4}c$ as a typical value
for the inner crust estimated by \cite{link_levin22}.  We fix the initial spin rate
to be 1 kHz (an unimportant choice).  

Assuming axisymmetry of the vorticity field, and neglecting vortex
motion in the azimuthal direction, the solution to the flow problem is
given entirely by eq. \ref{solution}.  As discussed at the end of \S
\ref{sv}, the effects of vortex tension are negligible at global
scales (but important at mesoscopic and microscopic scales) and we
neglect them.  The subsequent flow depends only on the spin rate of
the crust $\Omega(t)$ for specified $v_{\rm crit}(r)$. 

\begin{figure}[H]
\centering
\begin{tabular}{cccc}
\includegraphics[width=.33\linewidth]{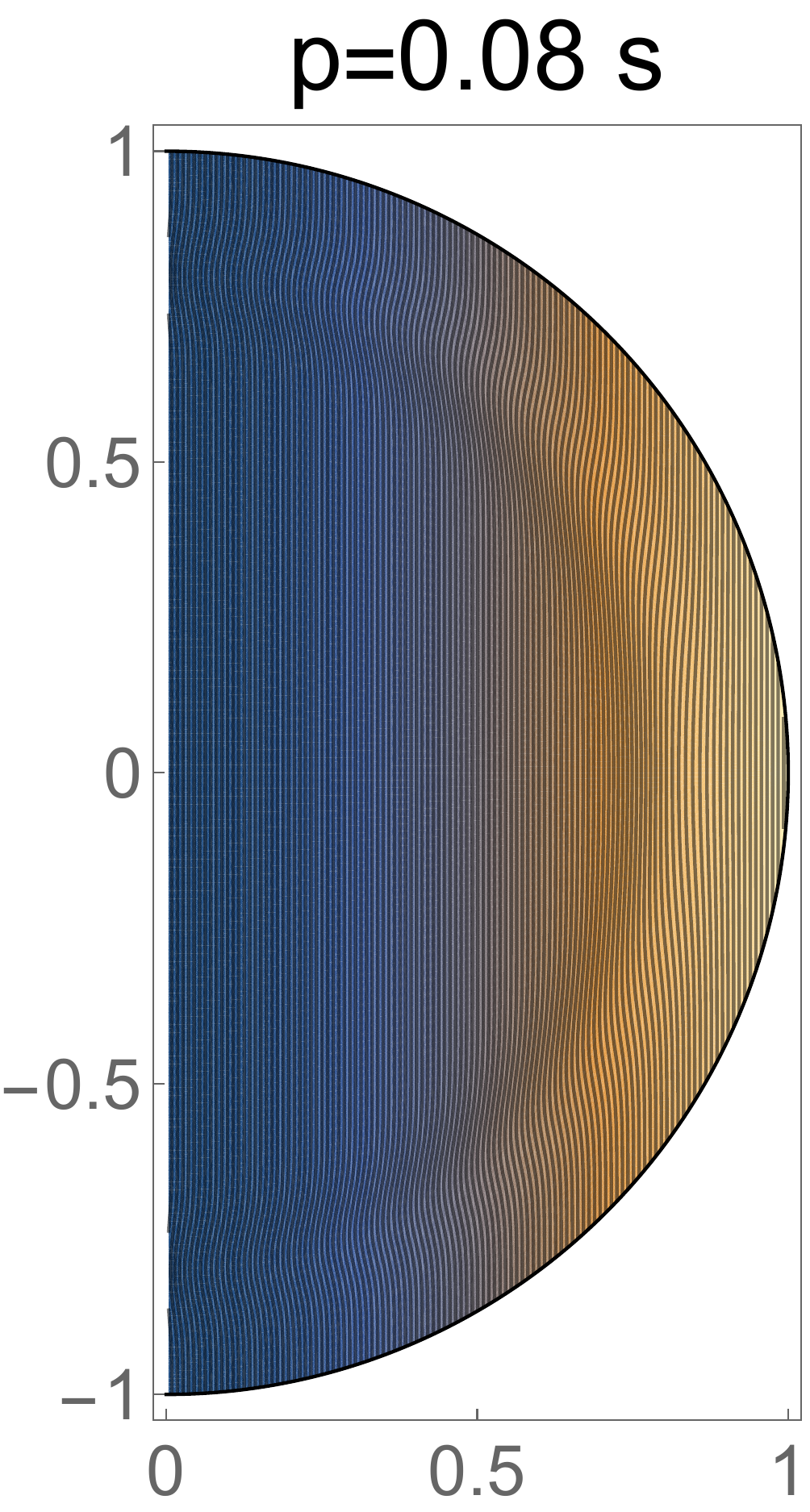} &
\includegraphics[width=.33\linewidth]{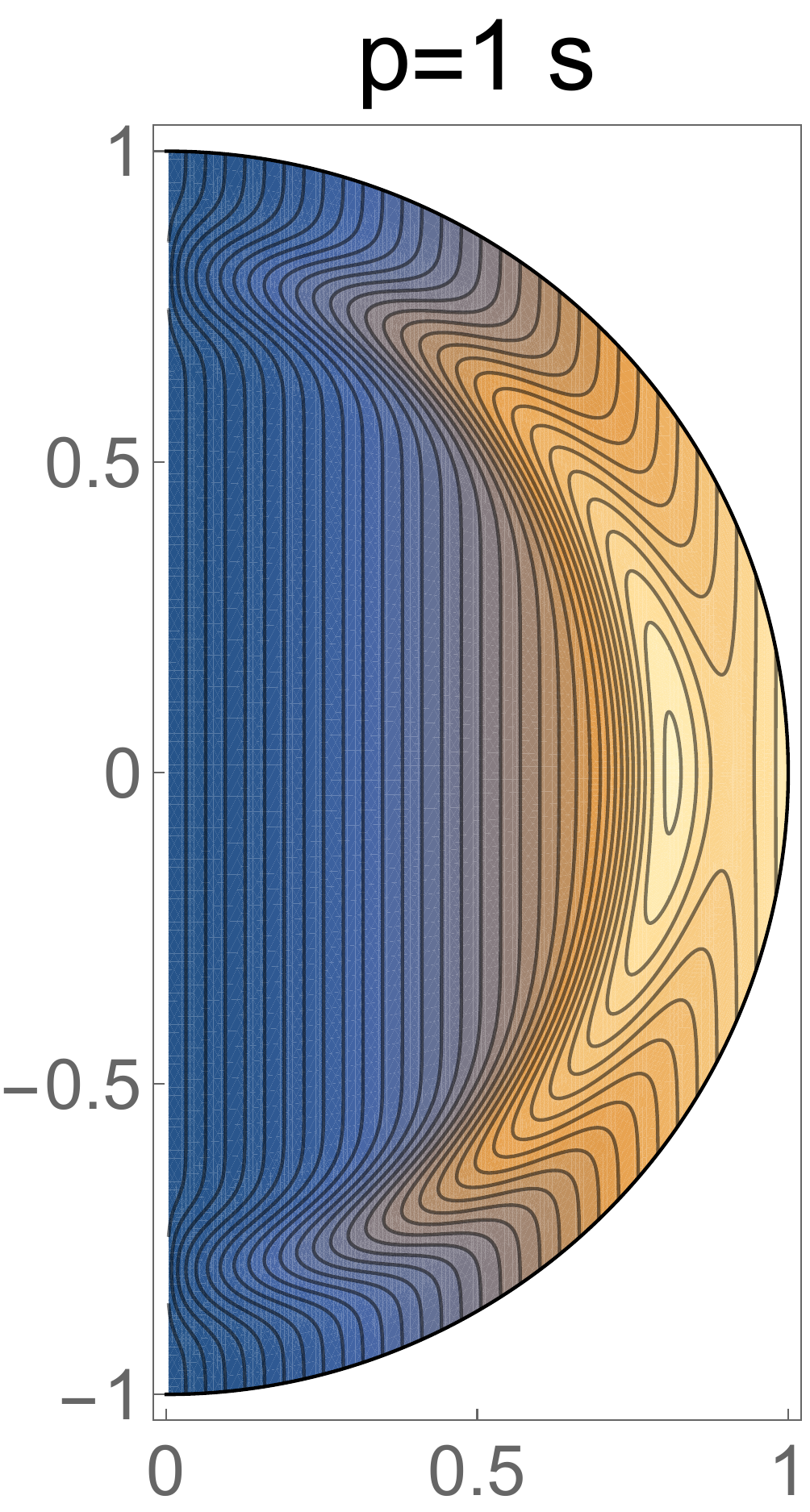} & 
\includegraphics[width=.33\linewidth]{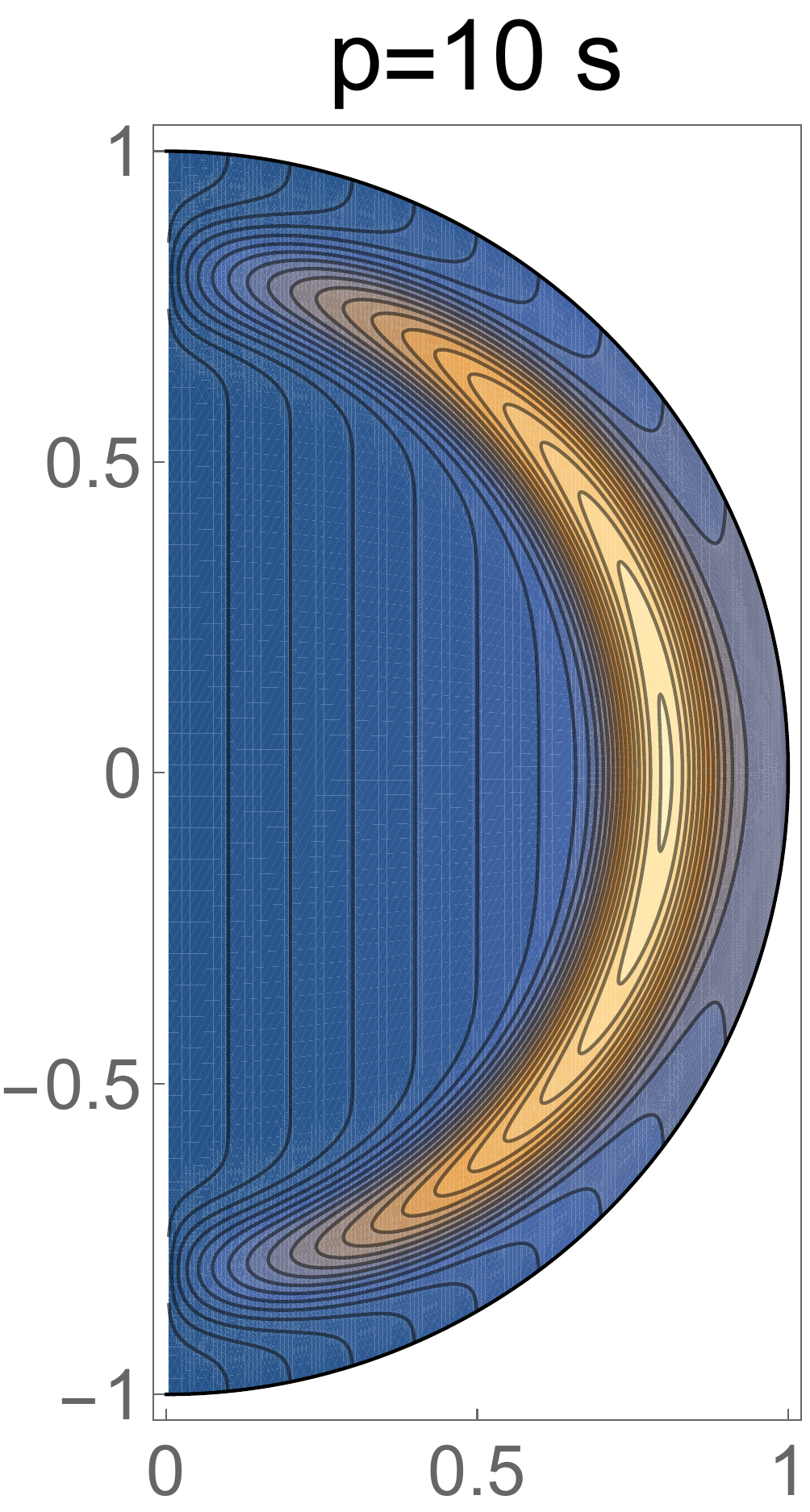} 
\end{tabular}
\caption{Contours of constant circulation for Model 1 for different
  spin rates. The spin axis
  is in the vertical direction. All dimensions are in units
of the NS radius, $\simeq 10$ km. The pinning region is centered on a
  spherical shell at $r=r_0=0.8$ with thickness $\sigma=0.05$.  }
\label{model-1}
\end{figure}

\begin{figure}[H]
\includegraphics[width=.2\linewidth]{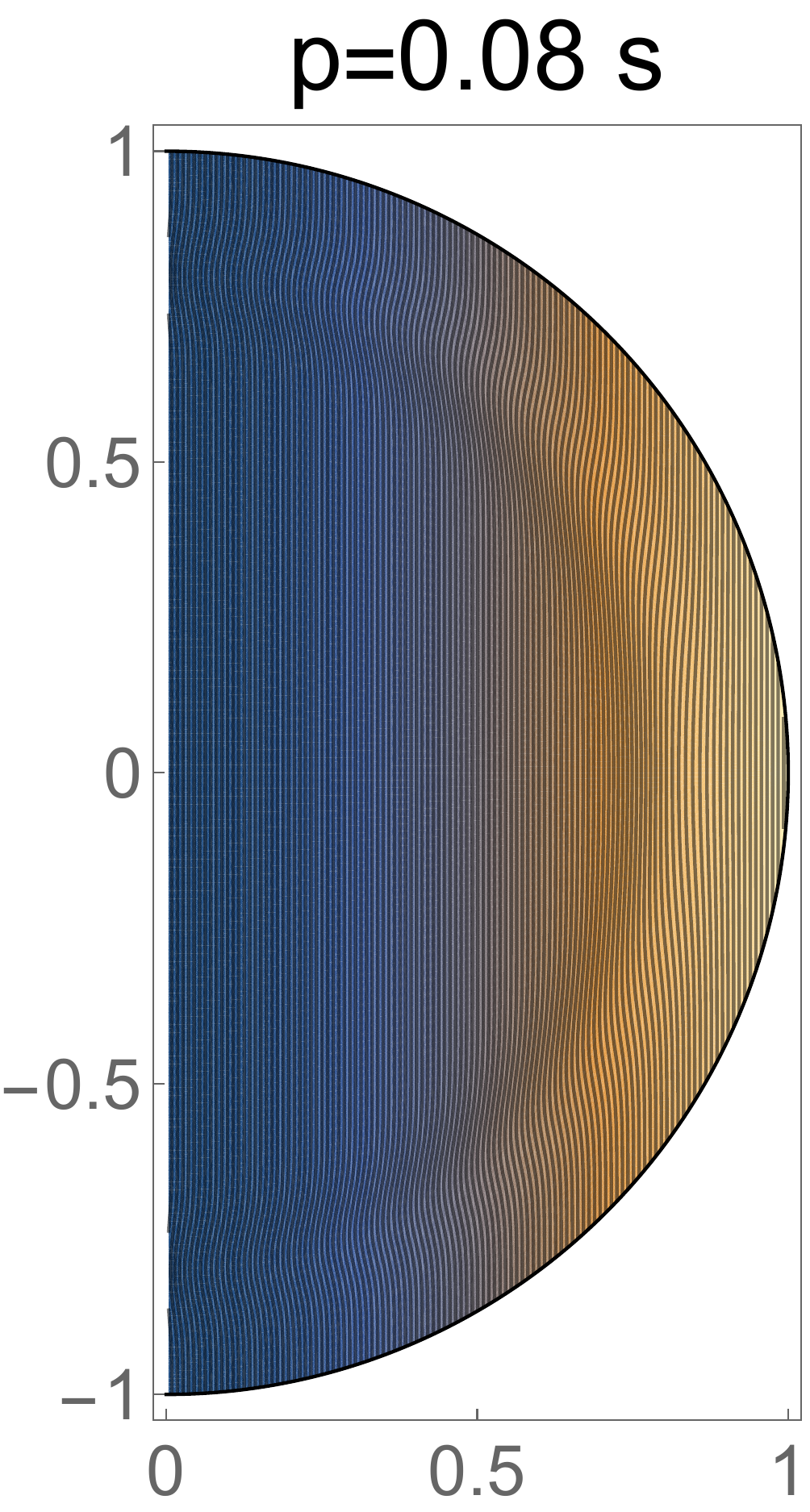}
\href{https://drive.google.com/file/d/1tuT1sUPG0MOggSQTu_TEsnpOdoN3gQNV/view?usp=share_link}{\\
\Large
  Click
  here to animate}
\caption{Animation of the simulation of Fig. \ref{model-1}.  Note the
  formation of tori within the shell that shrink and disappear. At
  late times (slow rotation), a dense torus of vortices exists
  that sustains a flow about the vertical axis in the equatorial
  plane. }
\label{m1}
\end{figure}

Fig. \ref{model-1} shows contours of constant circulation $\Phi $ at
different values of the crust spin rate for {\sl Example Model 1}.
These contours follow the vortex lines.  The contour interval is
chosen so that a uniform vortex distribution gives equally-spaced
contours.  All dimensions are in units of the NS radius $R\simeq 10$
km.  Fig. \ref{m1} shows an animation of the simulation of
Fig. \ref{model-1}.  As the star spins down, the vortex array is bent
into tori of vortex rings, and the vortices are compressed into dense
sheets at constant spherical radius.  Within the pinning  shell, vortex rings
shrink and disppear. At late times a dense torus of vortices remains
that sustains a flow - a superfluid ``river'' - about the vertical
axis in the equatorial plane.

\begin{figure}[H]
\centering
\begin{tabular}{cc}
\includegraphics[width=.45\linewidth]{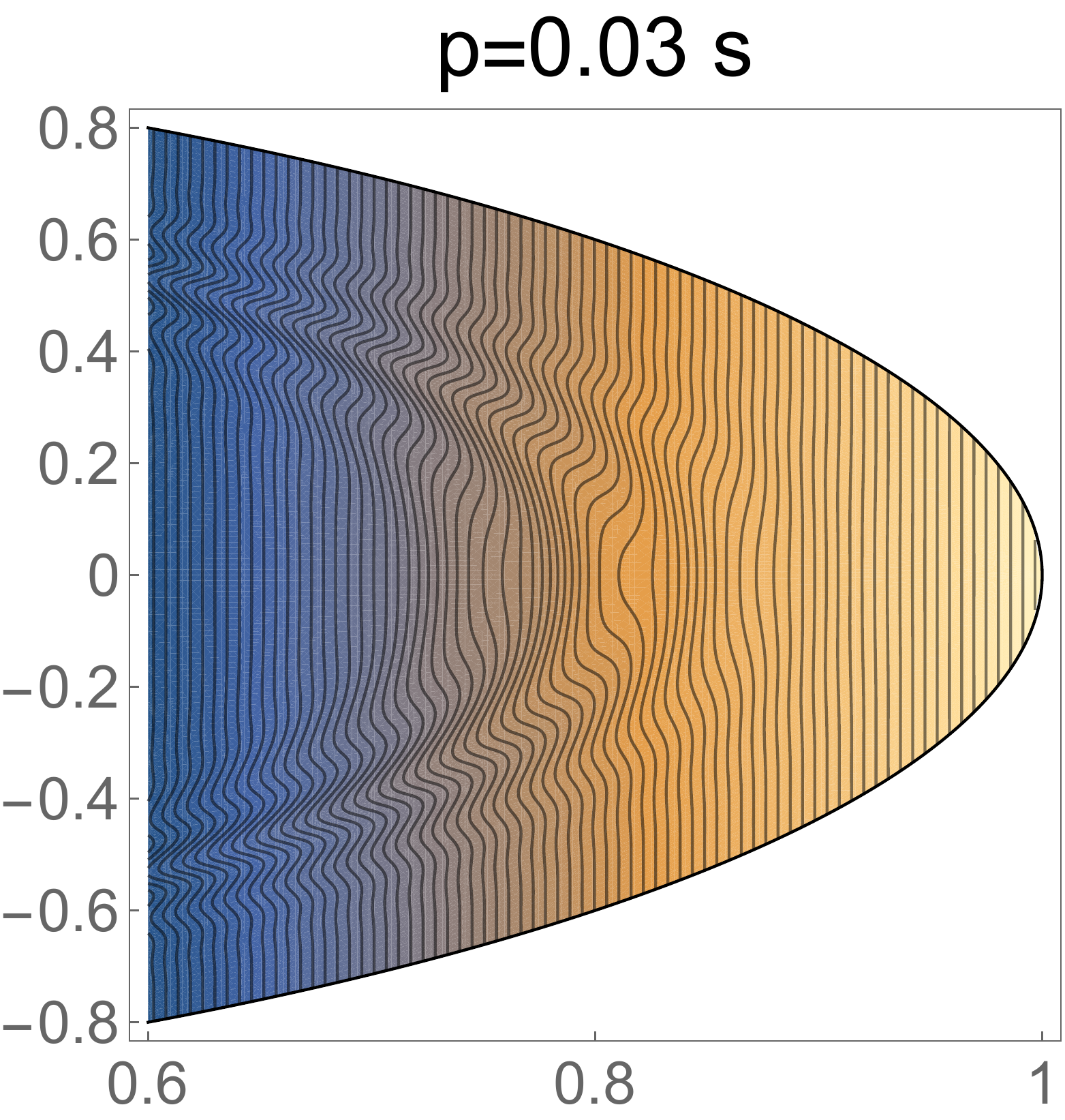} &
\includegraphics[width=.45\linewidth]{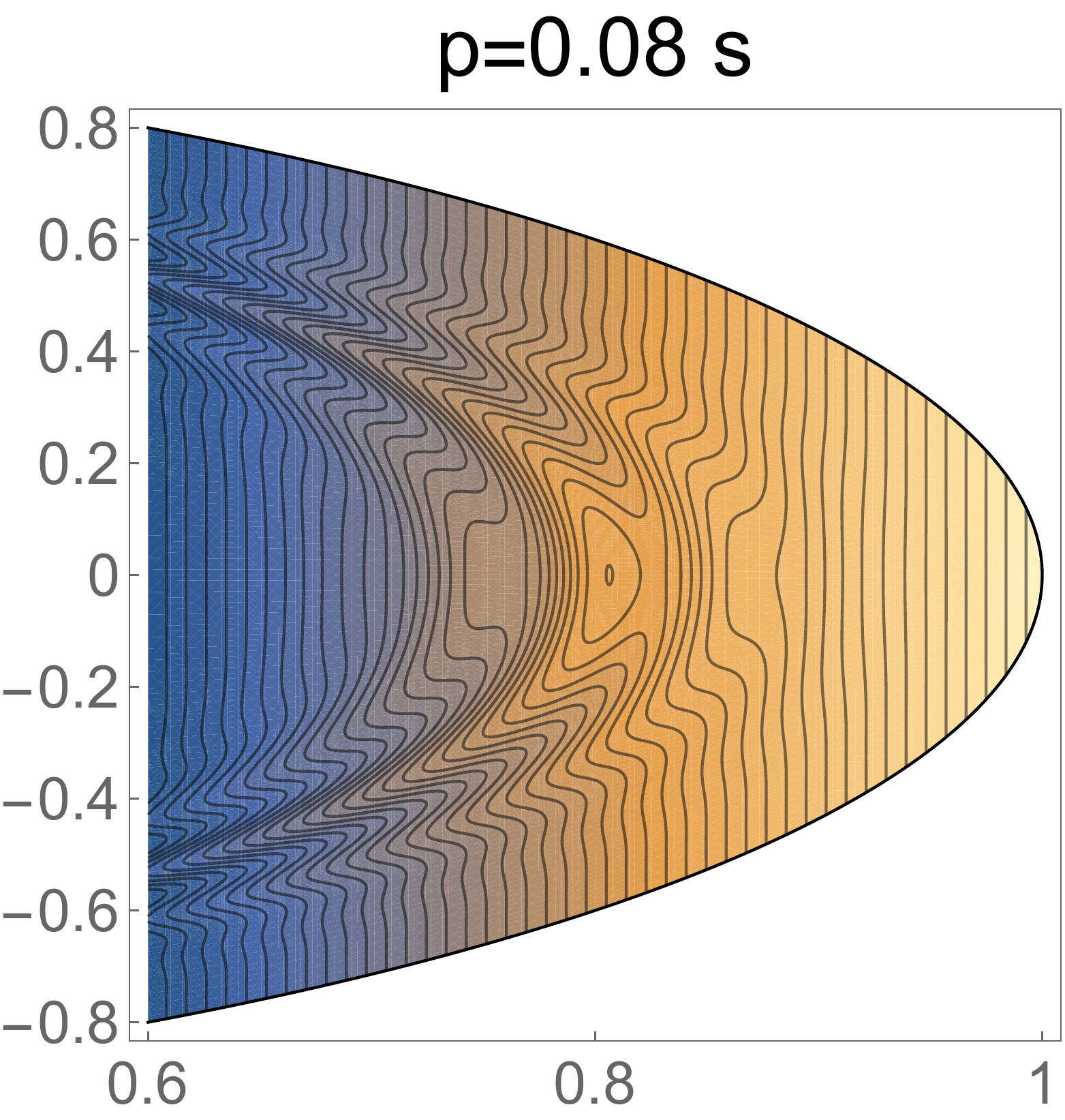} 
\end{tabular}
\hspace*{4pt}
\begin{tabular}{cc}
\includegraphics[width=.45\linewidth]{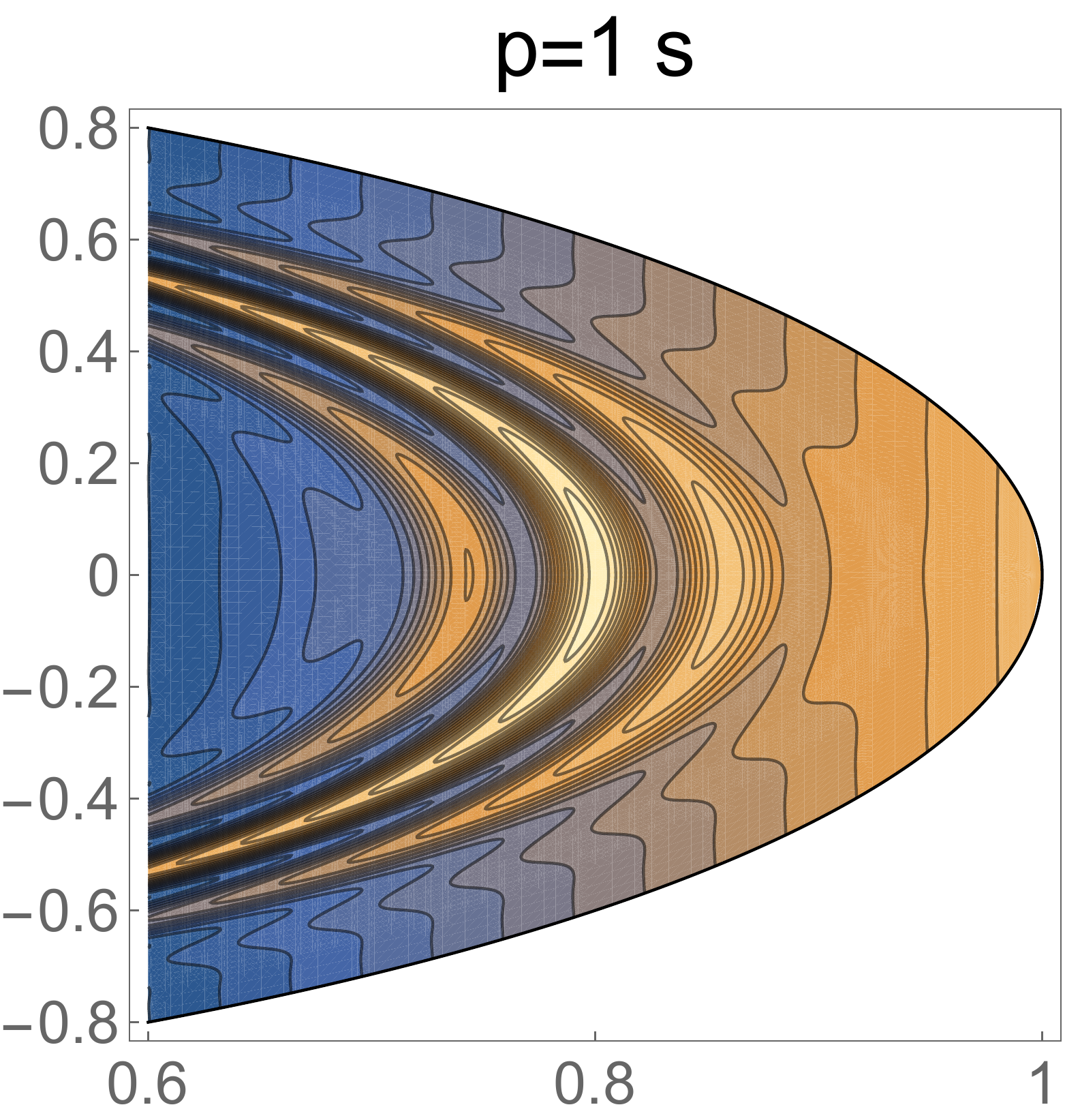} &
\includegraphics[width=.45\linewidth]{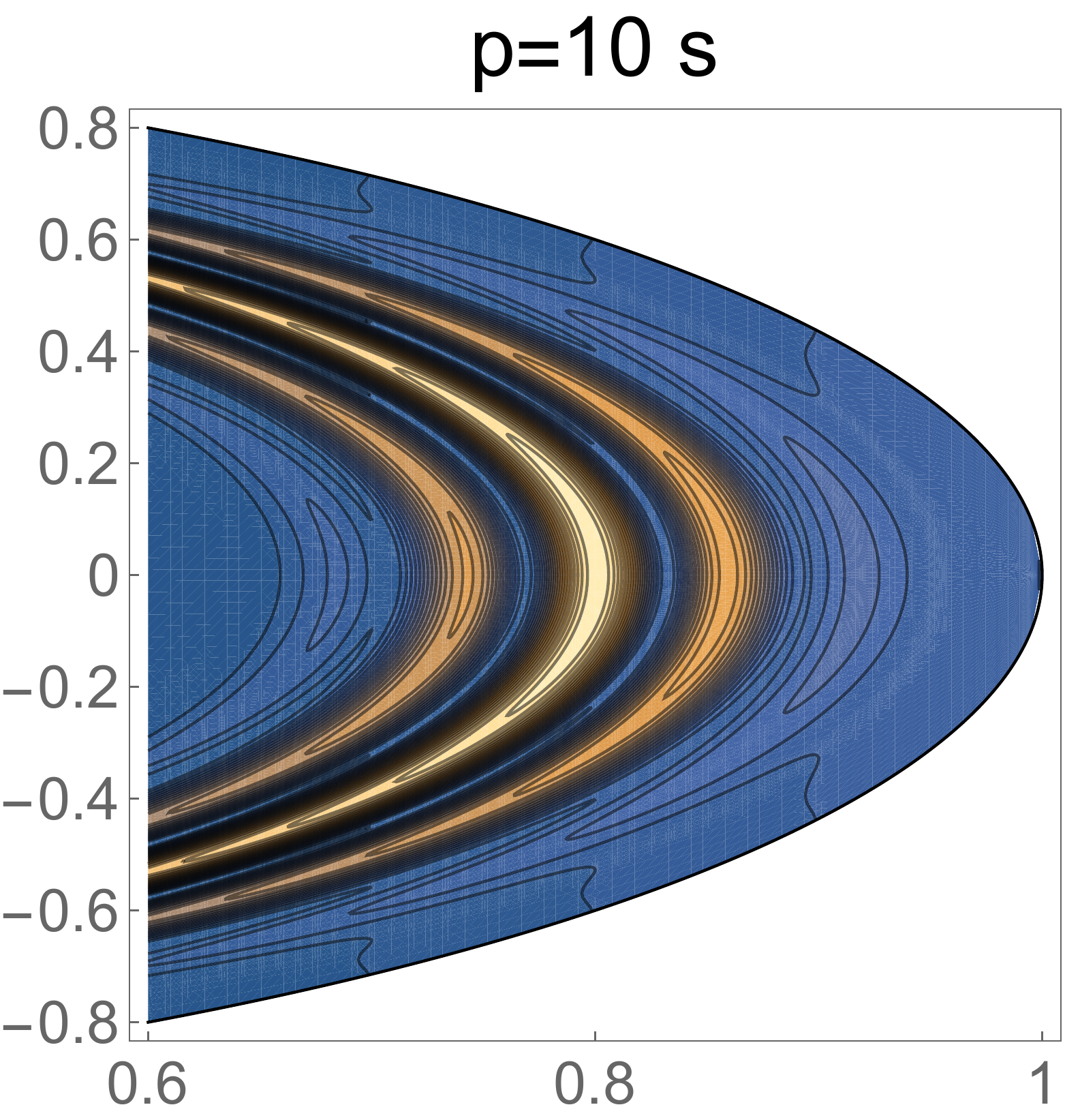} 
\end{tabular}
\caption{ Contours of constant circulation for Model 2 for different
  spin rates. The spin axis is in the vertical direction. The
  parameters are $r_0=0.8$, $\sigma=0.05$, $k=100$. The horizontal
  scale has been stretched. Note the formation of tori in multiple
  regions that shrink and disappear. At late times (slow rotation),
  distinct regions exist that sustain flow about the vertical axis in
  the equatorial plane. The three thin regions with light coloring in
  the two lower-most figures are nearly devoid of vortices. For this
  example, significant deformation of the vortex lattice exists even
  for a relatively short  spin period of 0.03 s (Crab). }
\label{model-2}
\end{figure}

\begin{figure}[H]
\includegraphics[width=.2\linewidth]{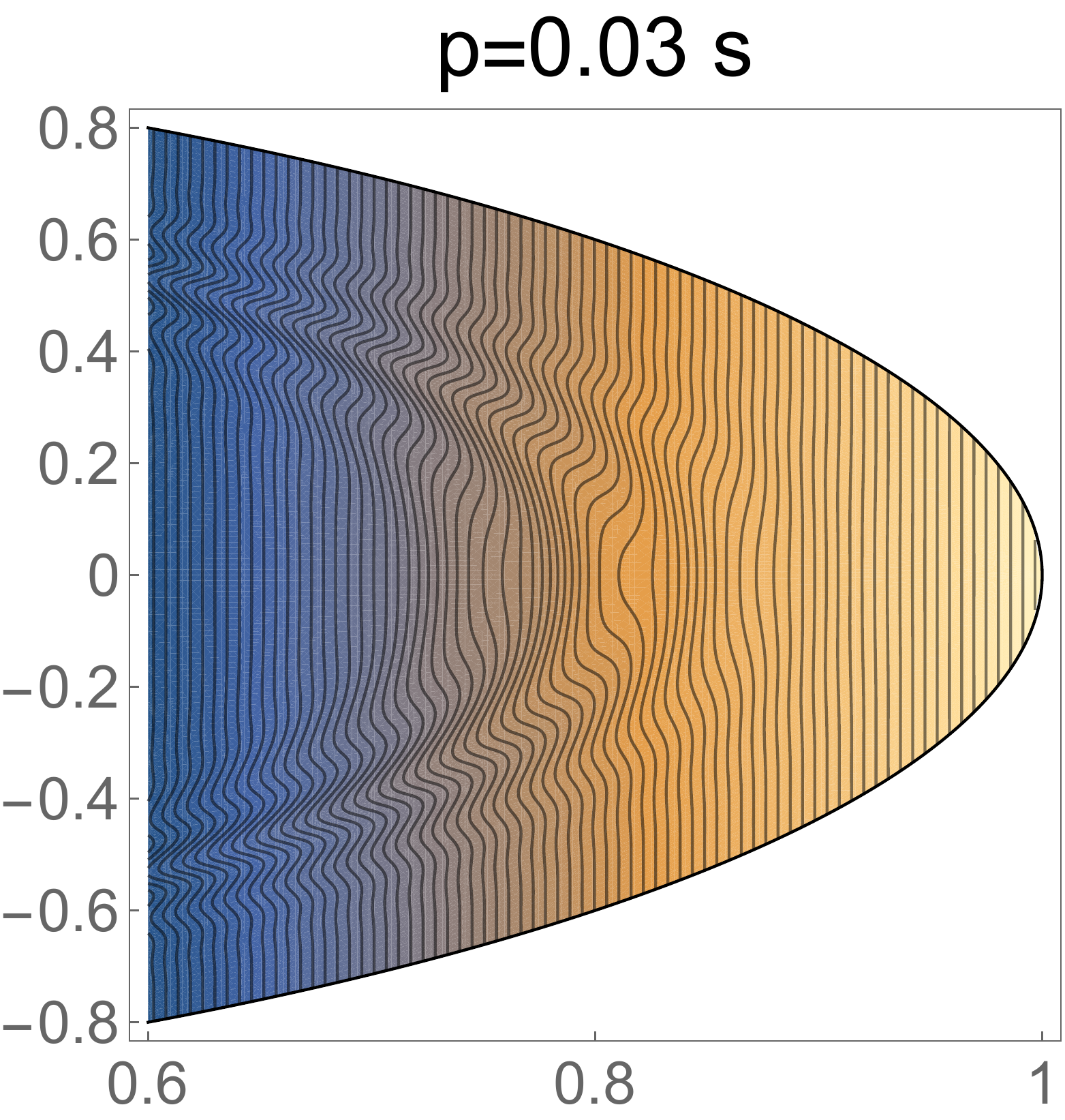}
\href{https://drive.google.com/file/d/1S9K5dxsIwNSLUqDujJByJt1BTpy14dC4/view?usp=share_link}{\\
  \Large Click
  here to animate}
\caption{Animation of the simulation of Fig. \ref{model-2}}
\label{m2}
\end{figure}

Fig. \ref{model-2} shows {\sl Example Model 2} with $k=100R^{-1}$,
giving three pinning maxima within a spherical shell.  An animation is
shown in Fig. \ref{m2}. The vortex array becomes significantly deformed
before a relatively short spin period of 0.03 s (the spin period
of the Crab pulsar) is reached.  Now vortices become compressed into multiple
sheets.  Shrinking tori are created in multiple locations. Bent
vortices at large polar radius connect to make shrinking tori while
their outer parts reconnect and exit the star.  At late times, the
vortex distribution freezes into multiple tori 
that carry flow around the rotation axis.  A modest
value of $k=100R^{-1}$ was selected to make easily readable plots;
pinning calculations indicate that configurations with stronger
gradients of $v_{\rm crit}$ are possible 
\citep{pvb97,donati_pizzochero2003,dp2004,dp06},  which would entail the
formation of vortex sheets at high rotation rates. 

\newpage

\newpage

\section{Discussion}
\label{discussion}

As a NS spins down, the
combined effects of the pinning force (which is always comparable to
the Coriolis force) and the Magnus force lead to 
significant bending of the vortex array. Bending of the vortex array
arises generally because pinning forces hold vortices largely stationary in
parts of the NS, while the Magnus force drives vortex motion where the
vortex array is not pinned. 

Despite the uncertainties of the distribution and magnitude of vortex
pinning in a NS, we can draw a number of robust conclusions.  First,
 the vortex
distribution will deviate significantly from rectilinear by the time
the spin rate goes below $\sim 20$ Hz (about twice the spin rate of
the Vela pulsar).  Later, the vortices become compressed into dense
sheets that follow surfaces of constant spherical radius.  The number
of sheets is equal to the number of radii in the star at which
$|d(rv_{\rm crit})/dr|$ is large.  If $d(rv_{\rm crit})/dr$ is first
large and positive at some radius, and then large and negative at
larger radius, the vortex distribution evolves into a torus that
spans the two regions. 

A region of particular interest is the base of the crust,
 where the nuclear lattice makes a transition from a lattice of
spherical nuclei to nuclear pasta, finally dissolving at the boundary
between the crust and the outer core.  Pinning forces must vary
significantly in this region, so we expect vortex sheets to be
present.  Dense sheets of opposing vorticity attract one another,
possibly leading to reconnection of vortex sheets.  If sheet reconnection
does occur, the superfluid flow could become turbulent at length scales
comparable to the typical sheet thickness. 

Our treatment depends on the assumption that the vortex array remains
poloidal in the rotating frame, thus avoiding the complications of
winding of the vortex array around the rotation axis of the star.
As discussed in \S \ref{sv}, calculations of vortex drag indicate
$\sin\theta\sim 10^{-3}$ in the inner crust and $\sim 0.1$ in the
outer core.  If vortices are forced by stellar spin-down to move in
the ballistic regime, there will be significant winding of the vortex
array, coupling the poloidal and toroidal motions of the array.  It is
more likely that vortices are never forced into the ballistic regime,
but instead move through thermally-activated vortex  creep. As we
show in a forthcoming publication \citep{link_levin2023}, for low-drag
thermal creep the vortex has comparable velocity components along and
transverse to $\vbf_s$, while for high-drag thermal creep the vortex
motion is nearly transverse to $\vbf_s^*$ as for zero temperature. 
We expect that the basic picture given here of the concentration of
vorticity into sense sheets  still holds, but with significant twisting
in the azimuthal direction. We will address this problem in a future
publication.

Our simplified treatment shows continuous evolution of a vortex
array. In reality, the vortex distribution might undergo abrupt
changes.  Abrupt changes
in the vortex distribution at large scales could be responsible for  spin
glitches, while smaller, stochastic changes could be related to timing
noise. 

YL's research on this subject is supported by Simons Investigator
Grant 827103. We thank Andrei Beloborodov, Ashley Bransgrove, Andrei
Gruzinov, and Malvin Ruderman for discussions.

\bibliography{references}{} \bibliographystyle{aasjournal}

\end{document}